
\magnification=1200
\baselineskip=13pt
\overfullrule=0pt
\tolerance=100000
\font\pq=cmr10 at 8truept

{\hfill \hbox{\vbox{\settabs 1\columns
\+ UR-1421 \cr
\+ ER-40685-870\cr
\+ hep-th/9505041\cr
}}}

\bigskip
\bigskip
\baselineskip=18pt

\centerline{\bf Bi-Hamiltonian Structure of the}
\centerline{\bf Supersymmetric Nonlinear Schr\"odinger Equation}
\vfill

\centerline{J. C. Brunelli}
\medskip
\centerline{and}
\medskip
\centerline{Ashok Das}
\medskip
\medskip
\centerline{Department of Physics and Astronomy}
\centerline{University of Rochester}
\centerline{Rochester, NY 14627, USA}
\vfill

\centerline{\bf {Abstract}}

\medskip
\medskip

We show that the supersymmetric nonlinear Schr\"odinger equation is a
bi-Hamiltonian integrable system. We obtain the two Hamiltonian structures
of the theory
from the ones of the supersymmetric two boson hierarchy through a field
redefinition. We also show how
the two Hamiltonian structures of the supersymmetric KdV equation can be
derived from a Hamiltonian reduction of the supersymmetric two boson hierarchy
as well.
\vfill
\eject

Bosonic integrable models have been studied in detail in the past [1-3].
These models have  a very rich structure. However, only recently,
after the discovery of the connection between hierarchies of
integrable equations and discretized versions of the two-dimensional gravity
[4], has there been a lot of interest in the high energy community in the
study of these systems. This also has led to a renewed interest in the study of
supersymmetric integrable systems for various reasons -- the
most important being
the fact that a supersymmetric theory of gravity would be free from problems
such as tachyonic states. However, a lot of properties of the supersymmetric
integrable systems remain to be studied.

The first supersymmetric integrable system to be studied was
the supersymmetric KP
hierarchy (sKP) of Manin and Radul [5], which upon appropriate reduction,
leads to the supersymmetric KdV equation (sKdV)
(the first fermionic extension of KdV, though, is due to Kupershmidt
[6]). The second Hamiltonian structure of
this system was shown by Mathieu [7] to
correspond to the superconformal algebra of the superstring theories. However,
the bi-Hamiltonian nature of the system was not known until much later. The
bi-Hamiltonian property is intimately connected with the
integrability of a system and this aspect of sKdV was obtained [8-10] from a
reduction of an even order sKP  Lax operator. It was found that the simplest
Hamiltonian structure of the KdV equation becomes a complicated nonlocal
structure upon supersymmetrization.

Another interesting supersymmetric integrable system that has received a lot of
attention, lately, is
the supersymmetric nonlinear Schr\"odinger equation (sNLS) [11-12].
In [11] it was shown that the most general supersymmetric
extension of the NLS is given by
$$
\eqalign{{
\partial  Q \over \partial  t} &=-(D^4 Q) + 2\alpha(D{\overline Q})(DQ)Q -
2\gamma{\overline Q}Q(D^2Q)+2(1-\alpha){\overline Q}(DQ)^2\cr
\noalign{\vskip 4pt}%
{\partial  \overline Q \over \partial  t} &= (D^4 \overline Q )-
2\alpha(DQ)(D{\overline Q}){\overline Q} +
2\gamma Q{\overline Q}(D^2{\overline Q})-2(1-\alpha)Q(D{\overline Q})^2\cr
}\eqno(1)
$$
where
$$
\eqalign{Q &= \psi + \theta q\cr
{\overline Q} &= {\overline\psi} + \theta {\overline q} \cr
}
\eqno(2)
$$
are fermionic superfields which are complex conjugates of each other, and
$$
D = {\partial  \over \partial  \theta} + \theta {\partial
 \over \partial  x}
\eqno(3)
$$
defines the supercovariant derivative in the superspace with coordinates
$z = (x,\theta)$
satisfying $D^2 = \partial$. However,
it was shown in [12] that the standard tests of integrability hold for the
system (1) only for
$$
\alpha=-\gamma=1\eqno(4)
$$
In other words, the supersymmetric extension of NLS which is also
integrable has the form,
$$
\eqalign{{
\partial  Q \over \partial  t} &=
 -(D^4 Q) + 2\left(D((DQ){\overline Q})\right)Q\cr
\noalign{\vskip 4pt}%
{\partial  \overline Q \over \partial  t} &=
 (D^4 \overline Q ) - 2\left(D((D{\overline Q})Q)\right){\overline Q}\cr
}\eqno(5)
$$
Much like the sKdV, it was shown in [12] that the naive supersymmetrization
of the Hamiltonian structures of the bosonic NLS equation goes through for the
second Hamiltonian structure only for the particular values of the parameters
in (4). However, a bi-Hamiltonian structure was still lacking. In this letter
we will derive the first Hamiltonian structure of the system showing that the
system is, indeed, bi-Hamiltonian for these values of the parameters.
This will complete the analysis of the integrable structure of
this system started in [12]. Like the sKdV system, the first structure, as we
will show, will be extremely nonlocal.

In ref. [13], it was shown that the scalar Lax operator for the sNLS system
can be identified with
$$
{\cal L}=-\left(D^2+{\overline Q}Q-{\overline Q}
D^{-1}(DQ)\right)\eqno(6)
$$
where $D^{-1}=\partial^{-1}D$ is the pseudo super-differential operator and
that the sNLS equations (5) can be  written as the nonstandard Lax equation
$$
{\partial{\cal L}\over\partial t}=\left[{\cal L},
\left({\cal L}^2\right)_{\ge1}\right]
\eqno(7)
$$
In principle, given a Lax equation, the Hamiltonian structures of the theory
can be derived from the Gelfand-Dikii brackets (appropriately extended to the
superspace) [3,10,14]. However, eq. (7) is a
nonstandard Lax equation and the definition
of the Gelfand-Dikii brackets have so far been extended only for the bosonic
systems in such cases [15]. The extension to superspace of such a
generalization
is technically much more involved and is
presently under study. However, here, we
will follow an alternate approach and exploit the relation between the
supersymmetric two boson system and the sNLS system to derive the Hamiltonian
structures for the latter.

The integrable supersymmetric two boson equation (sTB) is given by [16]
$$
\eqalign{{\partial  \Phi_0 \over \partial  t} &=
 - (D^4 \Phi_0) + (D(D\Phi_0)^2)+ 2(D^2 \Phi_1)\cr
\noalign{\vskip 4pt}%
{\partial  \Phi_1 \over \partial  t} &=
 (D^4 \Phi_1) + 2(D^2((D \Phi_0) \Phi_1))\cr
}
\eqno(8)
$$
where $\Phi_0$ and $\Phi_1$ are fermionic superfields. (We follow the notation
in ref. [16].)
This system can also be obtained from the following nonstandard Lax equation
$$
\eqalign{
L =& D^2 - (D \Phi_0) + D^{-1} \Phi_1\cr
{\partial{L}\over\partial t}=&\left[{L},
\left({L}^2\right)_{\ge1}\right]
}
\eqno(9)
$$
It has already been shown that (8) is a tri-Hamiltonian system [16,17]
$$
\partial_t\pmatrix{\Phi_0\cr
\noalign{\vskip 10pt}%
\Phi_1}
={\cal D}_1
\pmatrix{{\delta H_{3}\over\delta\Phi_0}\cr
\noalign{\vskip 10pt}%
{\delta H_{3}\over\delta\Phi_1}}=
{\cal D}_2
\pmatrix{{\delta H_{2}\over\delta\Phi_0}\cr
\noalign{\vskip 10pt}%
{\delta H_{2}\over\delta\Phi_1}}=
{\cal D}_3
\pmatrix{{\delta H_{1}\over\delta\Phi_0}\cr
\noalign{\vskip 10pt}%
{\delta H_{1}\over\delta\Phi_1}}
\eqno(10)
$$
where we note that the first two Hamiltonian structures have the form
$$
\eqalignno{
{\cal D}_1=&\pmatrix{0 & -D\cr
\noalign{\vskip 5pt}%
-D & 0}&(11a)\cr
\noalign{\vskip 15pt}%
{\cal D}_2=&\pmatrix{-2D-2D^{-1}\Phi_1D^{-1}+D^{-1}(D^2\Phi_0)D^{-1}&
D^3-D(D\Phi_0)+D^{-1}\Phi_1D\cr
\noalign{\vskip 20pt}%
-D^3-(D\Phi_0)D-D\Phi_1D^{-1}&-\Phi_1D^2-D^2\Phi_1}&(11b)
}
$$
The second Hamiltonian structure (the third as well) is highly nontrivial,
but it is important to note that it has been checked, using superprolongation
methods [18], that Jacobi identity holds for these structures [17].
The Hamiltonians  of this system are given by
$$
H_n={(-1)^{n+1}\over n}\hbox{sTr}\,L^n=
{(-1)^{n+1}\over n}\int dz\,\hbox{sRes}\,L^n\qquad n=1,2,\dots \eqno(12)
$$
where ``sRes'' is the super residue which is defined to be the
coefficient of the $D^{-1}$ term in the pseudo super-differential operator
with $D^{-1}$ at the right. The first few charges have the form
$$
\eqalign{
H_1=&-\int dz\, \Phi_1\cr
H_2=&-\int dz\, (D\Phi_0)\Phi_1\cr
H_3=&\int dz\,\Bigl[(D^3\Phi_0)-(D\Phi_1)-(D\Phi_0)^2\Bigr]\Phi_1\cr
H_4=&-{1\over2}\int
dz\,\Bigl[2(D^5\Phi_0)+2(D\Phi_0)^3+6(D\Phi_0)(D\Phi_1)-
3\left(D^2(D\Phi_0)^2\right)
\Bigr]\Phi_1\cr
}\eqno(13)
$$

The crucial observation [16], for our analysis, is the fact that
the sNLS equation and the sTB equation are
related to each other through the field redefinition (Miura transformation)
$$
\eqalignno{
\Phi_0 &= -\left(D \ln (DQ)\right) +
\left(D^{-1} (\overline Q Q)\right)&(14a)\cr
\Phi_1 &= - \overline Q ( DQ)
&(14b)
}
$$
The field redefinitions (14) allows us to write the Lax operator (9) also as
$$
L=G{\widetilde L}G^{-1}\eqno(15)
$$
where
$$
\eqalign{
G=&(DQ)^{-1}\cr
{\widetilde L}=&D^2-{\overline Q}Q-(DQ)D^{-1}{\overline Q} \cr
}\eqno(16)
$$
We say that $L$ and ${\widetilde L}$ are gauge related. All this is very much
like the bosonic case [19-22]. However, the important difference [13]
is that it is
not ${\widetilde L}$, rather its formal adjoint ${\widetilde L}^*$,
$$
{\widetilde L}^*=-\left(D^2+{\overline Q}Q-{\overline Q}
D^{-1}(DQ)\right)={\cal L}\eqno(17)
$$
which gives the sNLS equation as a nonstandard Lax equation (7). The relations
in (14) are invertible and can be formally written as
$$
\eqalignno{
Q=&\left(D^{-1}\hbox{e}^{\left(D^{-2}\left(-(D\Phi_0)+
\Phi_1(L^{-1}\Phi_0)\right)\right)}\right)
&(18a)\cr
{\overline Q}=&-\Phi_1
\hbox{e}^{\left(-D^{-2}\left(-(D\Phi_0)+\Phi_1(L^{-1}\Phi_0)\right)\right)}
&(18b)
}
$$
and these ((14) and (18)) define the connecting relation between the two
theories.

Given these relations between the two theories and given the fact that we
already know the Hamiltonian structures of the sTB system, we can obtain
the Hamiltonian structures of the sNLS equations in the following way. Let us
define the transformation matrix between the two systems as
$$
P = \left[{\delta(\Phi_0,\Phi_1)\over\delta(Q,{\overline Q})}\right]\eqno(19)
$$
where $\left[{\delta(\Phi_0,\Phi_1)\over\delta(Q,{\overline Q})}\right]$ is the
matrix formed from the Fr\'echet derivatives of $\Phi_0$ and $\Phi_1$ with
respect to $Q$ and ${\overline Q}$. It follows now that the Hamiltonian
structures of the two systems must be related as
$$
{\cal D}= P\,{\cal D}^{\,\hbox{\pq sNLS}} P^* \eqno(20)
$$
where $P^*$ is the formal adjoint of $P$ (with the matrix transposed). We note
here the explicit structure of the transformation matrix for completeness.
$$
P=\pmatrix{-D(DQ)^{-1}D+D^{-1}{\overline Q} & -D^{-1}Q\cr
\noalign{\vskip 10pt}%
-{\overline Q}D & -(DQ)}\eqno(21)
$$
It is useful to note that the matrix $P$ factorizes as
$$
P={\widetilde P}G\eqno(22)
$$
where
$$
\eqalignno{
{\widetilde P}=&\pmatrix{-D^{-1}&-D^{-1}(DQ)^{-1}Q D^2\cr
\noalign{\vskip 10pt}%
0&-D^2}&(23a)\cr
\noalign{\vskip 15pt}%
G=&\pmatrix{-{\widetilde L}^*(DQ)^{-1}D&0\cr
\noalign{\vskip 10pt}%
D^{-2}{\overline Q}D &D^{-2}(DQ)} &(23b)
}
$$
We also note from (20) that the Hamiltonian structures of the sNLS system
can be obtained from those of the sTB system (written in terms of $Q$ and
${\overline Q}$) as
$$
{\cal D}^{\,\hbox{\pq sNLS}}= P^{-1}{\cal D}\,(P^*)^{-1}\eqno(24)
$$
where the inverse matrix has the form
$$
P^{-1}=
\pmatrix{D^{-1}(DQ)({\widetilde L}^*)^{-1}D&
-D^{-1}(DQ)({\widetilde L}^*)^{-1}Q(DQ)^{-1}\cr
\noalign{\vskip 10pt}%
-{\overline Q}({\widetilde L}^*)^{-1}D &
-(1-{\overline Q}({\widetilde L}^*)^{-1}Q)(DQ)^{-1} }\eqno(25)
$$

Armed with these relations, we note that if we use the second Hamiltonian
structure of the sTB system, eq. (11b), we obtain from (24) that
$$
{\cal D}_2^{\,\hbox{\pq sNLS}} = P^{-1}{\cal D}_2(P^*)^{-1} \eqno(26)
$$
which after using (25) and a lot of tedious algebra gives
$$
{\cal D}_2^{\,\hbox{\pq sNLS}}=
\pmatrix{-QD^{-1}Q&-{1\over2}D+QD^{-1}{\overline Q}\cr
\noalign{\vskip 10pt}%
-{1\over2}D+{\overline Q}D^{-1}Q&
-{\overline Q}D^{-1}{\overline Q} }
\eqno(27)
$$
This is, indeed, the correct second Hamiltonian structure that was derived in
ref. [12] and provides a check on our method. (For those interested in
rederiving this result, we note that it is enormously simpler to check that
${\cal D}_2$ factorizes as
$$
{\cal D}_2 = P\,{\cal D}_2^{\,\hbox{\pq sNLS}}P^* \eqno(28)
$$
with ${\cal D}_2^{\,\hbox{\pq sNLS}}$ given in eq. (27).)

We are now in a position to derive the first Hamiltonian structure for the sNLS
system. We note from (24) that
$$
{\cal D}_1^{\,\hbox{\pq sNLS}} = P^{-1}{\cal D}_1(P^*)^{-1} \eqno(29)
$$
and using eqs. (11a) and (25), we obtain
$$
{\cal D}_1^{\,\hbox{\pq sNLS}}=
\pmatrix{
-D^{-1}(DQ)\Delta(DQ)D^{-1}
&
D^{-1}(DQ)\Delta{\overline Q}\cr
&
+D^{-1}(DQ)({\widetilde L}^*)^{-1}D^2(DQ)^{-1}\cr
\noalign{\vskip 30pt}%
{\overline Q}\Delta(DQ)D^{-1}
&
-{\overline Q}({\widetilde L}^*)^{-1}D^2(DQ)^{-1}-
(DQ)^{-1}D^2{\widetilde L}^{-1}{\overline Q}\cr
+(DQ)^{-1}D^2{\widetilde L}^{-1}(DQ)D^{-1}
&
-{\overline Q}\Delta{\overline Q}
}\eqno(30)
$$
where  we have defined
$$
\Delta\equiv
({\widetilde L}^*)^{-1}\left(D^2((DQ)^{-1}Q)\right){\widetilde L}^{-1}
\eqno(31)
$$
Like the first Hamiltonian structure of the sKdV equation, we note that this
structure is highly nonlocal and, therefore, could not be obtained from a naive
supersymmetrization of the corresponding bosonic structure. It can now be
checked explicitly that the sNLS equations (5) can be written in the
Hamiltonian form
$$
\partial_t\pmatrix{Q\cr
\noalign{\vskip 10pt}%
{\overline Q}}
={\cal D}_1^{\,\hbox{\pq sNLS}}
\pmatrix{{\delta H_{3}\over\delta Q}\cr
\noalign{\vskip 10pt}%
{\delta H_{3}\over\delta{\overline Q}}}=
{\cal D}_2^{\,\hbox{\pq sNLS}}
\pmatrix{{\delta H_{2}\over\delta Q}\cr
\noalign{\vskip 10pt}%
{\delta H_{2}\over\delta{\overline Q}}
}
\eqno(32)
$$
where the Hamiltonians can be obtained from (13) (and are defined in ref.
[13]).
(We note here that it is rather involved to check (32) for the first
structure in (30). It is much easier, however, to check that
$$
\pmatrix{{\delta H_{3}\over\delta Q}\cr
\noalign{\vskip 10pt}%
{\delta H_{3}\over\delta{\overline Q}}}=
\left({\cal D}_1^{\,\hbox{\pq sNLS}}\right)^{-1}
\pmatrix{ -(D^4 Q) + 2\left(D((DQ){\overline Q})\right)Q\cr
\noalign{\vskip 10pt}%
(D^4 \overline Q ) - 2\left(D((D{\overline Q})Q)\right){\overline Q}
}
\eqno(33)
$$
and the two are equivalent.) The two Hamiltonian structures satisfy the Jacobi
identity since the Hamiltonian structures of the sTB system do and define a
recursion operator which would relate all the Hamiltonian structures as well as
the conserved quantities of the system in a standard manner. However, as is
clear from the structure of the first Hamiltonian structure, it is extremely
nontrivial. Our result, therefore, shows that the
sNLS system (5) is a bi-Hamiltonian system and completes the analysis of the
integrability structure of this theory.

To conclude, we will now indicate how the Hamiltonian structures of the sKdV
equation
can be derived from those of the sTB system through a reduction. We note
that the sKdV [16] can be embedded into the sTB system in the following manner.
Let us look at the nonstandard Lax equation
$$
{\partial   L \over \partial  t} = [ L, (L^3)_{\geq 1}]
\eqno(34)
$$
with $L$ defined in (9). A simple calculation gives
$$
(L^3)_{\geq 1} = D^6 + 3 D \Phi_1 D^2 -3D^2(D\Phi_0)D^2+3(D\Phi_0)^2D^2
+6\Phi_1(D\Phi_0)D
\eqno(35)
$$
which leads to the dynamical equations (from (34))
$$
\eqalignno{
{\partial\Phi_1\over\partial  t} =&
-(D^6\Phi_1 ) - 3 D^2 \biggl( \Phi_1 (D \Phi_0)^2+(D^2\Phi_1)(D\Phi_0)+
\Phi_1(D\Phi_1)\biggr)&(36a)\cr
{\partial\Phi_0\over\partial  t} =&
\!-\!(D^6\Phi_0) + 3 D\biggl( \Phi_1 (D^2
\Phi_0)-2(D\Phi_1)(D\Phi_0)-\!{1\over3}(D\Phi_0)^3\!+\!
(D\Phi_0)(D^3\Phi_0)\!\biggr)&(36b)\cr
}
$$
We immediately see that the identification
$$
\eqalign{\Phi_0 & =  0\cr
\Phi_1 & = \Phi} \eqno(37)
$$
gives the sKdV equation
$$
{\partial\Phi\over\partial  t} =
-(D^6\Phi) - 3 D^2\left(\Phi(D\Phi)\right)\eqno(38)
$$
and shows how the sKdV equation is embedded in the sTB system as a nonstandard
Lax equation (34) with the Lax operator
$$
L = D^2 + D^{-1} \Phi\eqno(39)
$$
It is easy to see from (13) that with the condition in (37), the even
Hamiltonians vanish whereas the odd ones are the same as those for the sKdV
system.

The reduction in (37) imposes a constraint on the system. Consequently, the
Hamiltonian structures of the sKdV system can be obtained from those of the sTB
system through a Dirac procedure [23] as follows. Let ${\cal D}$ be one of the
Hamiltonian structures of the sTB system and $H$ denote one of the odd
Hamiltonians of the system. Then, we have
$$
\partial_t\pmatrix{\Phi_0\cr
\noalign{\vskip 10pt}%
\Phi_1}
={\cal D}
\pmatrix{{\delta H\over\delta\Phi_0}\cr
\noalign{\vskip 10pt}%
{\delta H\over\delta\Phi_1}}
\eqno(40)
$$
If we now use eq. (37) in (40), we obtain
$$
\partial_t\pmatrix{0\cr
\noalign{\vskip 10pt}%
\Phi}
=
\pmatrix{{\overline{\cal D}}_{11} & {\overline{\cal D}}_{12}\cr
\noalign{\vskip 5pt}%
{\overline{\cal D}}_{21}& {\overline{\cal D}}_{22}}
\pmatrix{{\overline{\delta H\over\delta\Phi_0}}=v_0\cr
\noalign{\vskip 10pt}%
{\overline{\delta H\over\delta\Phi_1}}=v_1}
\eqno(41)
$$
where $\overline{{\cal O}}$ denotes the quantities ${\cal O}$ calculated with
$\Phi_0=0$ and $\Phi_1=\Phi$. From (41), we immediately see that consistency
requires
$$
v_0=-{\overline{\cal D}}_{11}^{-1}{\overline{\cal D}}_{12}v_1\eqno(42)
$$
and that we can write
$$
\partial_t\Phi ={\overline{\cal D}}_{21}v_0 + {\overline{\cal D}}_{22}v_1
={\cal D}^{\,\hbox{\pq sKdV}}v_1 \eqno(43)
$$
with ($v_1$ for odd Hamiltonians is the same as ${{\delta
H}\over{\delta\Phi}}$ for the sKdV system.)
$$
{\cal D}^{\,\hbox{\pq sKdV}}={\overline{\cal D}}_{22}-
{\overline{\cal D}}_{21}{\overline{\cal D}}_{11}^{-1}{\overline{\cal D}}_{12}
\eqno(44)
$$
Using equation (11b), we note that
$$
({\overline{\cal D}}_2)^{-1}_{11}=-{1\over2}D(D^3+\Phi)^{-1}D\eqno(45)
$$
and using (44) we readily obtain the standard second Hamiltonian
structure of sKdV
$$
{\cal D}_2^{\,\hbox{\pq sKdV}}=-{1\over2}
(D^5+3\Phi D^2+(D\Phi)D+2(D^2\Phi))\eqno(46)
$$

The first Hamiltonian structure can also be obtained in a simple manner
through the Dirac reduction.
However, we should remember that since, in the limit $\Phi_0=0$, all the even
charges in (13) vanish, the first Hamiltonian structure of the sKdV is
obtained from ${\cal D}_0$ (and not from ${\cal D}_1$, as would be naively
expected) given by
$$
{\cal D}_0=R^{-1}{\cal D}_1\eqno(47)
$$
where $R={\cal D}_2{\cal D}_1^{-1}$ is the recursion operator of the sTB
system. It is easy to show that
$$
{\overline R}^{-1}=\pmatrix{4D\left({\cal D}_2^{\,\hbox{\pq
sKdV}}\right)^{-1}D&
-2D\left({\cal D}_2^{\,\hbox{\pq sKdV}}\right)^{-1}D^3\cr
\noalign{\vskip 20pt}%
2D^3\left({\cal D}_2^{\,\hbox{\pq sKdV}}\right)^{-1}D&
\qquad2D^2(D^3+\Phi)^{-1}D^{-1}(\Phi D^2+D^2\Phi)
\left({\cal D}_2^{\,\hbox{\pq sKdV}}\right)^{-1}D^3
}\eqno(48)
$$
Once again, using the Dirac reduction relation (44), we obtain
$$
{\cal D}_1^{\,\hbox{\pq sKdV}}=({\overline{\cal D}}_0)_{22}-
({\overline{\cal D}}_0)_{21}({\overline{\cal D}}_0)^{-1}_{11}
({\overline{\cal D}}_0)_{12}\eqno(49)
$$
This gives
$$
{\cal D}_1^{\,\hbox{\pq sKdV}}=-2D^2(D^3+\Phi)^{-1}D^2 \eqno(50)
$$
which is the nonlocal structure for the sKdV obtained in refs. [8-10]. Our
derivation, however, shows that this structure satisfies the Jacobi identity
since the Hamiltonian structures of sTB do. (The Jacobi identity for the first
Hamiltonian structure, to the best of our knowledge, has not yet been
demonstrated.)

To conclude, we have derived  in this letter, the bi-Hamiltonian structures of
the sNLS system starting from those of the sTB system. This completes the
analysis of the
integrability structure of the sNLS system. The derivation of these structures
as Gelfand-Dikii brackets remains an open question and is presently under
study. We have also shown how the Hamiltonian structures of the sKdV system can
be obtained from those of the sTB system through a Dirac reduction. This
provides an indirect proof of the Jacobi identity for these structures.
\bigskip
\noindent {\bf Acknowledgements}
\medskip

This work was supported in part by the U.S. Department of Energy Grant No.
DE-FG-02-91ER40685. J.C.B. would like to thank CNPq, Brazil, for
financial support.

\vfill\eject

\noindent {\bf {References}}
\bigskip

\item{1.} L.D. Faddeev and L.A. Takhtajan, ``Hamiltonian Methods in
the Theory of Solitons'' (Springer, Berlin, 1987).

\item{2.} A. Das, ``Integrable Models'' (World Scientific, Singapore,
1989).

\item{3.} L. A. Dickey, ``Soliton Equations and Hamiltonian Systems'' (World
Scientific, Singapore, 1991).

\item{4.} D. J. Gross and A. A. Midgal, Phys. Rev. Lett. {\bf 64}, 127 (1990);
D. J. Gross and A. A. Midgal, Nucl. Phys. {\bf B340}, 333 (1990); E. Br\'ezin
and V. A. Kazakov, Phys. Lett. {236B}, 144 (1990); M. Douglas and S. H.
Shenker,
Nucl. Phys. {\bf B335}, 635 (1990);
A. M. Polyakov in ``Fields, Strings and Critical Phenomena'', Les
Houches 1988, ed. E. Br\'ezin and J. Zinn-Justin (North-Holland, Amsterdam,
1989); L. Alvarez-Gaum\'e, Helv. Phys. Acta {\bf 64}, 361 (1991); P. Ginsparg
and G. Moore, ``Lectures on 2D String Theory and 2D Gravity'' (Cambridge, New
York, 1993).

\item{5.} Y. I. Manin and A. O. Radul, Commun. Math. Phys. {\bf 98}, 65 (1985).

\item{6.} B.A. Kupershmidt, Phys. Lett. {\bf A102}, 213 (1984).

\item{7.} P. Mathieu, J. Math. Phys. {\bf 29}, 2499 (1988).

\item{8.} W. Oevel and Z. Popowicz, Comm. Math. Phys. {\bf 139}, 441 (1991).

\item{9.} J.M. Figueroa-O'Farril, J. Mas and E. Ramos, Leuven preprint
KUL-TF-91/19 (1991); J. M. Figueroa-O'Farrill, J. Mas and E. Ramos,
Rev. Math. Phys. {\bf 3}, 479 (1991).

\item{10.} J. Barcelos-Neto and A. Das, J. Math. Phys. {\bf 33}, 2743 (1992).

\item{11.} G. H. M. Roelofs and P. H. M. Kersten, J. Math. Phys. {\bf 33}, 2185
(1992).

\item{12.} J. C. Brunelli and A. Das, J. Math. Phys. {\bf 36}, 268 (1995).

\item{13.} J. C. Brunelli and A. Das, ``A Nonstandard Supersymmetric KP
Hierarchy'', University of Rochester preprint UR-1367 (1994) (also
hep-th/9408049), to appear in the Rev. Math. Phys..

\item{14.} A. Das and W.-J. Huang, J. Math. Phys. {\bf 33}, 2487 (1992).

\item{15.} J. C. Brunelli, A. Das and W.-J. Huang, Mod. Phys. Lett. {\bf 9A},
2147 (1994).

\item{16.} J. C. Brunelli and A. Das, Phys. Lett. {\bf B337}, 303 (1994).

\item{17.} J. C. Brunelli and A. Das, ``Properties of Nonlocal Charges in the
Supersymmetric Two Boson Hierarchy'', University of Rochester
preprint UR-1417 (1995) (also hep-th/9504030).

\item{18.} P. Mathieu, Lett. Math. Phys. {\bf 16}, 199 (1988).

\item{19.} H. Aratyn, L.A. Ferreira, J.F. Gomes and A.H. Zimerman, Nucl.
Phys. {\bf B402}, 85 (1993); H. Aratyn, L.A. Ferreira, J.F. Gomes and A.H.
Zimerman, ``On $W_\infty$ Algebras, Gauge Equivalence of KP Hierarchies,
Two-Boson Realizations and their KdV Reductions'', in Lectures at the VII
J. A. Swieca Summer School, S\~ao Paulo, Brazil, January 1993,
eds. O. J. P. \'Eboli and V. O. Rivelles (World Scientific, Singapore, 1994);
H. Aratyn, E. Nissimov and S. Pacheva, Phys. Lett. {\bf B314}, 41 (1993).

\item{20.} L. Bonora and C.S. Xiong, Phys. Lett. {\bf B285}, 191 (1992); L.
Bonora and C.S. Xiong, Int. J. Mod. Phys. {\bf A8}, 2973 (1993).

\item{21.} M. Freeman and P. West, Phys. Lett. {\bf 295B}, 59 (1992).

\item{22.} J. Schiff, ``The Nonlinear Schr\"odinger Equation and
Conserved Quantities in the Deformed Parafermion and SL(2,{\bf R})/U(1)
Coset Models'', Princeton preprint IASSNS-HEP-92/57 (1992)
(also hep-th/9210029).

\item{23.} W. Oevel and O. Ragnisco, Physica {\bf A161}, 181 (1989).
\end